\documentclass[a4paper]{article}
\usepackage{amsmath}
\usepackage{amssymb}
\usepackage{pstricks}
\usepackage{epsfig}
\usepackage{graphicx}
\usepackage{color}
\usepackage{cite}
\usepackage{mathrsfs}

\begin{document}
\begin{center}
{\large \sc \bf On the dispersionless Davey-Stewartson hierarchy: Zakharov-Shabat equations, twistor structure and Lax-Sato formalism }

\vskip 20pt

{\large G. Yi  }

\vskip 20pt

{\it
 School of Mathematics, Hefei University of Technology,\\
 Hefei Anhui 230601, China \\
}

\bigskip

e-mail:  {\tt ge.yi@hfut.edu.cn}

\bigskip

\end{center}

\begin{abstract}

In this paper, we continue the study of the Davey-Stewartson system which is one of the most important$(2+1)$ dimensional integrable models. As we showed in the previous paper, the dDS (dispersionless Davey-Stewartson) system arises from Hamiltonian vector fields Lax pair. A new hierarchy of compatible PDEs defining infinitely many symmetries, which is associated with the dDS system, is defined in this paper. We show that this hierarchy arises from the commutation condition of a particular series of one-parameter Hamiltonian vector fields.

\end{abstract}

\section{Introduction}
In 1974 Davey and Stewartson \cite{DavSte} used a multi-scale analysis to derive a coupled system of nonlinear partial differential equations describing the evolution of a three dimensional wave packet in water of a finite depth. The general DS(Davey-Stewartson) system (parametrized by $\varepsilon >0$) for a complex (wave-amplitude) field $q(x,y,t)$ and a real (mean-flow) field $\phi(x,y,t)$  is given by
\begin{subequations} \label{DavSte}
\begin{eqnarray}
\textbf{i}\varepsilon q_{t}+\frac{\varepsilon^{2}}{2}(q_{xx}+\sigma^{2} q_{yy})+\delta q \phi=0,
\end{eqnarray}
\begin{eqnarray}
\sigma^{2}\phi_{yy}-\phi_{xx}+(|q|^{2})_{xx}+\sigma^{2}(|q|^{2})_{yy}=0,
\end{eqnarray}
\end{subequations}
in which the variables $t,x,y \in \mathbb{R}$. We shall refer to (\ref{DavSte}) with $\sigma=1$ the DS-I (Davey-Stewartson-I) system and with $\sigma=\textbf{i}$ the DS-II (Davey-Stewartson-II) system. We also refer to (\ref{DavSte}) with $\delta=1$ the focusing case and  with $\delta=-1$ the defocusing case respectively.
The Davey-Stewartson system (\ref{DavSte}), as a prototype example of classical integrable systems, have been extensively studied and many important results are obtained \cite{AblCla}: $N$-line soliton solutions \cite{AnkFre,APPa,APPb,Naka1,Naka2,SatAbl}; localized exponentially decaying solitons \cite{BLMP}; an infinite dimensional symmetry group,in fact this involves an infinite dimensional Lie algebra with a Kac-Moody-Virasoro loop structure\cite{ChaWin,Omote}; similarity reductions to the second and fourth Painlev\'{e} equations \cite{Tajiri} ; a B\"{a}cklund transformation and Painlev\'{e} property\cite{GanLak}; an infinite number of commuting symmetries and conservation quantities, a recursion operator and bi-Hamiltonian structure\cite{FokSan,SanFok}.

Konopelchenko, Taimanov studied the infinitely many symmetries of DS system and pointed out that any symmetry induces an infinite family of geometrically different deformations of tori in $\mathbb{R}^{4}$ preserving the Willmore functional. They defined the DS hierarchy by considering the compatibility of undetermined differential operators in terms of $\partial_{z}$ and $\hat{\partial}_{z}$ \cite{Kono1,Kono2,Taimanov} and gave examples of $t_{2}$ and $t_{3}$ flows.

Although it is well known that the infinite many symmetries of DS system are closely connected to the 2-component KP hierarchy \cite{konopelmulti,AblCla}, however the expression of the multi-component KP hierarchy is usually formulated in a matrix form with matrix pseudo-differential operators. These complex definitions and calculations yield technical difficulties when one tries to study the higher symmetry flows of DS system in detail. Recently, we gave a simple and clear expression for the DS hierarchy by two scalar pseudo-differential operators involved with $\partial,\hat{\partial}$ and proved the compatibility for these infinite flows \cite{Yi1}.

Konopelchenko proposed the quasiclassical generalized Weierstrass representation (GWR) for highly
corrugated surfaces in $\mathbb{R}^{4}$  with a slow modulation. They pointed out that the integrable
deformations of such surfaces are described by the dDS  hierarchy \cite{konopelchenko2007}. Takasaki and Takebe identified the universal Whitham hierarchy of genus zero with a multi-component KP hierarchy and they pointed out the difficulties for defining the reasonable dispersionless limitation since the matrix operators and vector wave functions\cite{tt2007}. Therefore, similar to the classical case, it is difficult to study the higher symmetry flows of dDS system in detail. The main result of this paper is to define a new and clear expression for dDS hierarchy based on two pairs of eigenfunctions of a specific vector field and to show its structure including Hamiltonians.

In the recent research \cite{yi2018}, we showed that, the following dDS system, which is the semi-classical limit of Davey-Stewartson system,
\begin{subequations} \label{dDS system}
\begin{eqnarray}
u_{t}+2(u S_{z})_{z}-2(u S_{\hat{z}})_{\hat{z}}=0,
\end{eqnarray}
\begin{eqnarray}
S_{t}+S^{2}_{z}-S^{2}_{\hat{z}}+\phi=0,
\end{eqnarray}
\begin{eqnarray}
\phi_{z\hat{z}}+2(u_{zz}-u_{\hat{z}\hat{z}})=0,
\end{eqnarray}
\end{subequations}
arises from the commutation condition of Hamiltonian vector fields Lax pair
\begin{eqnarray} \label{cccc}
\left[L_{1},L_{2}\right]=0
\end{eqnarray}
in which
\begin{subequations} \label{L12}
\begin{eqnarray}
L_{1}=\partial_{\hat{z}}-\{H_{1},\cdot\},
\end{eqnarray}
\begin{eqnarray}
L_{2}=\partial_{t}-\{H_{2},\cdot\},
\end{eqnarray}
\end{subequations}
and  the  Hamiltonians read as follows
\begin{subequations} \label{H12}
\begin{eqnarray}
H_{1}=S_{\hat{z}}+\frac{u}{\lambda},
\end{eqnarray}
\begin{eqnarray} \label{H2}
H_{2}=(\lambda^{2}-2S_{z} \lambda)+\left[\left(S^{2}_{z}-2 \partial^{-1}_{z} (u_{\hat{z}})\right)+\frac{2u S_{\hat{z}}}{\lambda}+\frac{u^{2}}{\lambda^{2}}\right].
\end{eqnarray}
\end{subequations}
Here and hereafter in this paper, the Poisson bracket is defined as $\{A,B\}=A_{\lambda}B_{z}-A_{z}B_{\lambda}$. Obviously, this dDS system (\ref{dDS system}) is equivalent to the following Zakharov-Shabat equation
\begin{eqnarray}
\frac{\partial H_{1}}{\partial t}-\frac{\partial H_{2}}{\partial \hat{z}}+\{H_{1},H_{2}\}=0.
\end{eqnarray}

An integrable PDE (partial differential equation) system is usually associated with a hierarchy of PDEs defining infinitely many symmetries. It is also the case for the integrable dispersionless (hydrodynamic type) systems. Two important examples the dKP (dispersionless Kadomtsev-Petviashvili) hierarchy and SDIFF(2) Toda hierarchy which have been intensively studied in the literatures \cite{krichever,kodama,KG,manakov2,manakov3,takebe1,takasaki1,takasaki2,takasaki3,takasaki4,takasaki5}. The dispersionless Davey-Stewartson equation and hierarchy have been first introduced and discussed by Konopelchenko \cite{konopelchenko2007}. The structures and properties of the dispersionless Davey-Stewartson hierarchy will be discussed in more detail in this paper.

According to (\ref{H12}), in this framework, $0$ and $\infty$ are the singular points in the complex $\lambda$-plane. Therefore the construction of dDS hierarchy is different from the dKP case which has the only singular point $\infty$ in the complex $\lambda$-plane. In the dDS case, the generators of the Hamiltonians are two eigenfunctions, one with polar singularity around $0$ and the other around $\infty$. In fact, this case is closely connected with the framework of hyper-K\"{a}hler hierarchy developed by Takasaki  \cite{takebe1,takasaki1,takasaki2,takasaki3,takasaki4,takasaki5} but  more general since both eigenfunctions appear in the construction of the Hamiltonians (i.e., including both the positive and negative parts).

In fact, the uniform expression (\ref{dDS system}) includes both the dDS-I and dDS-II system.  By considering the change of independent variables $z=x+\sigma y, \hat{z}=\textbf{i}(x-\sigma y)$ and the change of dependent variables
$u(z,\hat{z},t)=\textbf{i}\tilde{u}(x,y,t),S(z,\hat{z},t)=\tilde{S}(x,y,t),\phi(z,\hat{z},t)=\tilde{\phi}(x,y,t)$, the original expression (\ref{dDS system}) reads as follows

\begin{subequations} \label{dDSreal}
\begin{eqnarray}
\tilde{u}_{t}+(\tilde{u} \tilde{S}_{x})_{x}+(\tilde{u} \tilde{S}_{y})_{y}=0,
\end{eqnarray}
\begin{eqnarray}
\tilde{S}_{t}+\frac{1}{2}(\tilde{S}^{2}_{x}+\sigma^{2} \tilde{S}^{2}_{y})+ \phi=0,
\end{eqnarray}
\begin{eqnarray}
4\tilde{u}_{xx}-\tilde{\phi}_{xx}+\sigma^{2} (4\tilde{u}_{yy}+\tilde{\phi}_{yy})=0,
\end{eqnarray}
\end{subequations}
in which $\sigma=1$ corresponds the dDS-I system and $\sigma=\textbf{i}$ corresponds the dDS-II system. In this expression (\ref{dDSreal}), the real-valued independent variables $x,y$  are spatial variables and $t$ is the time variable, they all possess the definite physical meanings. Hereafter in this paper, we can always consider this change of independent variables $z=x+\sigma y, \hat{z}=\textbf{i}(x-\sigma y)$ when we study the whole hierarchy.

This paper is organized as follows: first, we define the dDS (dispersionless Davey-Stewartson) hierarchy which includes infinitely many compatible PDEs; second, we derive the twistor structure and Lax-Sato formlism for this dDS hierarchy; third, we show some examples including dDS system, complex dmKP (dispersionless modified KP) system and discuss the more general unreduced hierarchy.

\section{The dispersionless Davey-Stewartson hierarchy}
The notion of dDS hierarchy is an extension of the system (\ref{dDS system}) associated to the vector fields Lax pair (\ref{L12}) and Hamiltonians (\ref{H12}). From this point we introduce a vector filed as
\begin{eqnarray} \label{hat L}
\hat{L}&=&\partial_{\hat{z}}-\{\hat{H},\cdot\}=\partial_{\hat{z}}-\{v+\frac{u}{\lambda},\cdot\}  \nonumber\\
&=& \partial_{\hat{z}}+\frac{u}{\lambda^{2}} \partial_{z}+(v_{z}+\frac{u_{z}}{\lambda}) \partial_{\lambda},
\end{eqnarray}
where $\lambda$ is a complex parameter and  $u=u(t_{mn}),v=v(t_{mn})$ depend on complex variables $t_{mn}$ ($m,n$ are nonnegative integers and $m+n \geq 1$, and we denote $t_{10} \equiv z,t_{01} \equiv \hat{z}$ in this paper.)

With a given arbitrary closed curve $\Gamma$ around the original point  in the complex $\lambda$-plane, two eigenfunctions  $\Phi$ and $\hat{\Phi}$ of $\hat{L}$  can be described by
the following formal Laurent expansions:
\begin{subequations}
\begin{eqnarray} \label{Phi}
\Phi&=&\lambda+\sum_{k \leq 0}f_{k} \lambda^{k},
\end{eqnarray}
\begin{eqnarray} \label{hatPhi}
\hat{\Phi}&=&\frac{u}{\lambda}+\sum_{k \geq 0}g_{k} \lambda^{k}.
\end{eqnarray}
\end{subequations}
The eigenfunction $\Phi$  is defined and holomorphic outside $\Gamma$ ( include this curve) up to the point $\lambda=\infty$ where it has a first order pole, and respectively
 the eigenfunction $\hat{\Phi}$ is defined and holomorphic inside $\Gamma$ ( include this curve) up to the point $\lambda=0$ where it has a first order pole.

\bigskip
\bigskip
\bigskip
\textbf{Remark1}: By recursion, we can deduce that all the coefficients $f_{k}$ and $g_{k}$ of the Laurent expansions  $\Phi$ and $\hat{\Phi}$ depend on the two functions $u,v$ and their derivatives or integrals with respect to the independent variables $z,\hat{z}$.
 The leading terms of the eigenfunctions $\Phi$ and $\hat{\Phi}$ can be obtained directly by the definition $\hat{L} \Phi=0$ and $\hat{L} \hat{\Phi}=0$:
\begin{eqnarray}
f_{0}&=&-\partial^{-1}_{\hat{z}}(v_{z}),~~~f_{-1}=-\partial^{-1}_{\hat{z}}(u_{z}),~~f_{-2}=\partial^{-1}_{\hat{z}}[u
\partial^{-1}_{\hat{z}}(v_{zz})-v_{z}\partial^{-1}_{\hat{z}}(u_{z})],\cdot\cdot\cdot  \nonumber \\
g_{0}&=&v,~~~~~~~~g_{1}=-\frac{\partial^{-1}_{z}(u_{\hat{z}})}{u},~~~~~~~~~~g_{2}=\frac{\partial^{-1}_{z}[v_{z}\partial^{-1}_{z}
(u_{\hat{z}})-u v_{\hat{z}}]}{u^{2}},\cdot\cdot\cdot
\end{eqnarray}
Then all the coefficients $f^{(m)}_{k}$ and $g^{(n)}_{k}$ of the following Laurent expansions
\begin{subequations}
\begin{eqnarray}
\Phi^{m}&=&\lambda^{m}+\sum_{k \leq m-1}f^{(m)}_{k} \lambda^{k},
\end{eqnarray}
\begin{eqnarray}
\hat{\Phi}^{n}&=&\frac{u^{n}}{\lambda^{n}}+\sum_{k \geq 1-n}g^{(n)}_{k} \lambda^{k},
\end{eqnarray}
\end{subequations}
also depend on $u,v$ and their derivatives or integrals with respect to the independent variables $z,\hat{z}$.

\bigskip
\bigskip
\bigskip
\textbf{Remark2}: According to the expression (\ref{Phi}), the parameter $\lambda$ can be expressed formally in terms of Laurent expansions in $\Phi$
\begin{eqnarray} \label{lambda1}
\lambda=\Phi+\sum_{k \leq 0} p_{k} \Phi^{k},
\end{eqnarray}
in which $p_{k}=p_{k}(f_{0},f_{-1},\cdots,f_{k}).$
 Similarly, the Hamiltonian $\hat{H}=v+\frac{u}{\lambda}$ can also be expressed formally in terms of Laurent expansions in$\hat{\Phi}$
\begin{eqnarray} \label{hatH1}
\hat{H}=\hat{\Phi}+\sum_{k \leq -1} q_{k} \hat{\Phi}^{k},
\end{eqnarray}
in which $q_{k}=q_{k}(u,g_{0},g_{1},\cdots,g_{-k}).$  Therefore, all the coefficients $p_{k},q_{k}$ from the above two Laurent expansions depend on $u,v$ and their derivatives or integrals with respect to the independent variables $z,\hat{z}$.

\bigskip
According to (\ref{H2}), the second Hamiltonian associated with the dDS system (\ref{dDS system}) can be expressed as
$(\Phi^{2})_{>0}+(\hat{\Phi}^{2})_{\leq 0}$. Here and hereafter $()_{>0}$ stands for the extracting the positive  powers of $\lambda$ and the similar definition for $()_{\leq 0}$. This notion can be generalized to arbitrary nonnegative integers $m,n$ by introducing the Hamiltonians as follows
\begin{eqnarray} \label{Hmn}
&&H_{mn}=\mathscr{A}_{m}+\mathscr{B}_{n},\nonumber \\
&&H_{m0}=\mathscr{A}_{m},~~H_{0n}=\mathscr{B}_{n},
\end{eqnarray}
in which
\begin{eqnarray}
&&\mathscr{A}_{m}=(\Phi^{m})_{>0}, \nonumber\\
&&\mathscr{B}_{n}=(\hat{\Phi}^{n})_{\leq 0}.
\end{eqnarray}

\bigskip
\textbf{Proposition 1.} The Zakharov-Shabat equations
\begin{eqnarray} \label{simplezerocurvature}
\frac{\partial \hat{H}}{\partial t_{mn}}-\frac{\partial H_{mn}}{\partial \hat{z}}+\{\hat{H},H_{mn}\}=0
\end{eqnarray}
are equivalent to the following $(2+1)$ dimensional closed systems for unknowns $u$ and $v$:
\begin{subequations} \label{simplesystems}
\begin{eqnarray}
u_{t_{mn}}-(u f^{(m)}_{1})_{z}-g^{(n)}_{-1,\hat{z}}=0,
\end{eqnarray}
\begin{eqnarray}
v_{t_{mn}}+f^{(m)}_{0,\hat{z}}-g^{(n)}_{0,\hat{z}}=0.
\end{eqnarray}
\end{subequations}

\bigskip
In fact, due to the ring properties of the space of eigenfunctions of vector fields, also $\Phi^{m}$ and
$\hat{\Phi}^{n}$ satisfy the vector field equation $\hat{L} (\Phi^{n})=0,\hat{L} (\hat{\Phi}^{n})=0$, from which one
infers the recursive relations:
\begin{subequations} \label{recur}
\begin{eqnarray}
&&f^{(m)}_{k,\hat{z}}+u f^{(m)}_{k+2,z}+(k+1)v_{z}f^{(m)}_{k+1}+(k+2)u_{z} f^{(m)}_{k+2}=0,\nonumber\\
&&f^{(m)}_{m}=1,~~~~~f^{(m)}_{k}=0~~~~~~(k >m)
\end{eqnarray}
\begin{eqnarray}
&&g^{(n)}_{k,\hat{z}}+u g^{(n)}_{k+2,z}+(k+1)v_{z}g^{(n)}_{k+1}+(k+2)u_{z} g^{(n)}_{k+2}=0,\nonumber\\
&&g^{(n)}_{-n}=u^{n},~~~~~g^{(n)}_{-k}=0~~~~~~(k >n)
\end{eqnarray}
\end{subequations}

The Zakharov-Shabat equations (\ref{simplezerocurvature}) are equivalent to
\begin{eqnarray} \label{HtmnExpression}
\frac{\partial \hat{H}}{\partial t_{mn}}&=&\hat{L} (H_{mn})\nonumber\\
&=&\hat{L}\left[(\Phi^{m})_{>0}\right]+\hat{L}\left[(\hat{\Phi}^{n})_{\leq 0}\right]\nonumber\\
&=& v_{t_{mn}}+\frac{u_{t_{mn}}}{\lambda}
\end{eqnarray}
Substituting the recursive relations (\ref{recur})  into the above expression (\ref{HtmnExpression}), one obtains the systems (\ref{simplesystems}).

The notion of dDS (dispersionless Davey-Stewartson) is a straightforward extension of the above special case (which include the particular Hamiltonian $\hat{H}$).

\bigskip
\textbf{Definition 1.} The Zakharov-Shabat equations
\begin{eqnarray} \label{zerocurvature}
\frac{\partial H_{mn}}{\partial t_{kl}}-\frac{\partial H_{kl}}{\partial t_{mn}}+\{H_{mn},H_{kl}\}=0
\end{eqnarray}
are equivalent to a hierarchy of compatible systems. These compatible systems are defined as the dDS (dispersionless Davey-Stewartson) hierarchy.

\bigskip
In fact, the Zakharov-Shabat equations (\ref{simplezerocurvature}) are equivalent to the following commutation conditions
\begin{eqnarray} \label{commu1}
\left[\hat{L},L_{mn}\right]=0,
\end{eqnarray}
in which the vector fields Lax pairs are defined as
\begin{subequations}
\begin{eqnarray}
\hat{L}=\partial_{\hat{z}}-\{\hat{H},\cdot\}=\partial_{\hat{z}}-\frac{\partial \hat{H}}{\partial \lambda} \partial_{z}+ \frac{\partial \hat{H}}{\partial z} \partial_{\lambda},
\end{eqnarray}
\begin{eqnarray}
L_{mn}=\partial_{t_{mn}}-\{H_{mn},\cdot\}=\partial_{t_{mn}}-\frac{\partial H_{mn}}{\partial \lambda} \partial_{z}+ \frac{\partial H_{mn}}{\partial z} \partial_{\lambda}.
\end{eqnarray}
\end{subequations}
Therefore, all the vector fields $L_{mn} (m+n \geq 2)$ share the same eigenfunctions with $\hat{L}$. This fact implies
\begin{eqnarray} \label{commu1}
\left[L_{mn},L_{kl}\right]=0,
\end{eqnarray}
which is equivalent to the Zakharov-Shabat equations (\ref{zerocurvature}). Therefore the Zakharov-Shabat equations (\ref{zerocurvature}) are equivalent to compatible systems.

\bigskip
Analogous to the structure of dKP hierarchy and SDIFF(2) Toda hierarchy\cite{takebe1,takasaki1}, the Zakharov-Shabat equations (\ref{zerocurvature}) take place in the analysis of the self-dual vacuum Einstein equation and hyper-K\"{a}hler geometry \cite{takasaki6}. In the study of the vacuum Einstein equation (as well as its hyper-K\"{a}hler version), a K\"{a}hler-like 2-form and associated "Darboux coordinates" play a central role \cite{gindikin,hitchin}. The dDS hierarchy has another expression which resembles the twistor structure.

\textbf{Definition 2.} By introducing an exterior differential  2-form as
\begin{eqnarray} \label{omega}
\omega=\sum_{m+n \geq 1} dH_{mn}\wedge dt_{mn}=d \lambda \wedge d z+d \hat{H} \wedge d \hat{z}+\sum_{m+n \geq 2} dH_{mn}\wedge dt_{mn},
\end{eqnarray}
then the dDS hierarchy can be defined as
\begin{eqnarray} \label{2nddef}
\omega=d\Phi\wedge d\Psi=d\hat{\Phi}\wedge d\hat{\Psi},
\end{eqnarray}
in which
\begin{subequations} \label{eigen}
\begin{eqnarray}
\Phi&=&\lambda+\sum_{k \leq 0}f_{k} \lambda^{k},
\end{eqnarray}
\begin{eqnarray} \label{Psi}
\Psi=z+\sum_{m+n \geq 2} m t_{mn} \Phi^{m-1} +\sum_{j \leq -2} v_{j} \Phi^{j},
\end{eqnarray}
\begin{eqnarray}
\hat{\Phi}&=&\frac{u}{\lambda}+\sum_{k \geq 0}g_{k} \lambda^{k},
\end{eqnarray}
\begin{eqnarray} \label{hatPsi}
\hat{\Psi}=\hat{z}+\sum_{m+n \geq 2} n t_{mn} \hat{\Phi}^{n-1} +\sum_{j \leq -2} \hat{v}_{j} \hat{\Phi}^{j}.
\end{eqnarray}
\end{subequations}
The symbol $d$  stands for total differentiation in $\lambda,z,\hat{z}$ and $t_{mn}~(m+n \geq 2)$.

\bigskip
\bigskip
In fact, firstly, $\omega$ is a closed form, i.e.,
\begin{eqnarray}
d\omega=0.
\end{eqnarray}
Secondly, the Zakharov-Shabat equations (\ref{zerocurvature}), which give the definition of dDS hierarchy, can be cast into a compact form as
\begin{eqnarray}
\omega \wedge \omega=0.
\end{eqnarray}
These two relations imply the existence of two functions $L$ and $M$ that give a pair of Darboux coordinates as
\begin{eqnarray}
\omega=d L \wedge d M.
\end{eqnarray}
More specifically, one may select $L=\Phi$ which is defined in (\ref{Phi}) and to find the formal solution for the equation
 \begin{eqnarray} \label{omegaPsi}
 \omega=d \Phi \wedge d\Psi.
\end{eqnarray}
This equation can be rewritten as
\begin{eqnarray} \label{partial1}
d \left(\Psi d\Phi+\lambda d z+\hat{H} d \hat{z}+\sum_{m+n \geq 2} H_{mn}d t_{mn}\right)=0.
\end{eqnarray}
This implies the existence of a function $F=F(\Phi,z,\hat{z},t_{mn})$ such that
\begin{eqnarray}
d F=\Psi d\Phi+\lambda d z+\hat{H} d \hat{z}+\sum_{m+n \geq 2} H_{mn}d t_{mn},
\end{eqnarray}
which is  equivalent to
\begin{eqnarray}
\frac{\partial F}{\partial \Phi}&=&\Psi,~~~~\frac{\partial F}{\partial z}=\lambda,\nonumber\\
\frac{\partial F}{\hat{z}}&=&\hat{H},~~~~~~~\frac{\partial F}{\partial t_{mn}}=H_{mn}.
\end{eqnarray}
By considering $H_{mn}$'s  positive part (positive powers of $\lambda$), one obtains the expression of $F$ as follows
\begin{eqnarray} \label{F}
F=z \Phi+\sum_{m+n \geq 2} t_{mn} \Phi^{m}+\sum_{j \leq 0} \theta_{j} \Phi^{j},
\end{eqnarray}
Since the relation (\ref{lambda1}), the nonpositive part of  $F$ in the expression (\ref{F}) is expressed by the powers of $\Phi$  and
all the coefficients $\theta_{j} (j \leq 0)$ depend on $u,v$ and their derivatives or integrals with respect to the independent variables $z,\hat{z}$.
By direct calculation, one obtains
\begin{eqnarray}
\Psi=z+\sum_{m+n \geq 2} m t_{mn} \Phi^{m-1} +\sum_{j \leq -2} v_{j} \Phi^{j},
\end{eqnarray}
 where $v_{j}=(j+1)\theta_{j+1}$. Therefore the pair $(\Phi,\Psi)$ forms a Darboux coordinate of $\omega$.

 Analogous to the above process, a function $\hat{F}=\hat{F}(\hat{\Phi},z,\hat{z},t_{mn})$ can be constructed as the following form
\begin{eqnarray} \label{hatF}
\hat{F}=\hat{z} \hat{\Phi}+\sum_{m+n \geq 2} t_{mn} \hat{\Phi}^{n}+\sum_{j \leq -1} \hat{\theta}_{j} \hat{\Phi}^{j},
\end{eqnarray}
which satisfies
\begin{eqnarray}
d \hat{F}=\hat{\Psi} d\hat{\Phi}+\lambda d z+\hat{H} d \hat{z}+\sum_{m+n \geq 2} H_{mn}d t_{mn},
\end{eqnarray}
By direct calculation, one obtains
\begin{eqnarray}
\hat{\Psi}=\hat{z}+\sum_{m+n \geq 2} n t_{mn} \hat{\Phi}^{n-1} +\sum_{j \leq -2} \hat{v}_{j} \hat{\Phi}^{j}.
\end{eqnarray}
where $\hat{v}_{j}=(j+1)\hat{\theta}_{j+1}$. Therefore the pair $(\hat{\Phi},\hat{\Psi})$ also forms a Darboux coordinate of $\omega$.

\bigskip
\textbf{Definition 3.} The dDS (dispersionless Davey-Stewartson) hierarchy consists of the following Lax-Sato equations
\begin{eqnarray} \label{LS}
\frac{\partial K}{\partial t_{mn}}=\{H_{mn},K\}
\end{eqnarray}
for eigenfunctions $K=\Phi,\Psi,\hat{\Phi},\hat{\Psi}$  defined in (\ref{eigen}) and the canonical Poisson relations
\begin{eqnarray} \label{canonical}
\{\Phi,\Psi\}=1,~~~~\{\hat{\Phi},\hat{\Psi}\}=1.
\end{eqnarray}
These Lax-Sato equations are equivalent to the commutation conditions

\begin{eqnarray}
[L_{mn},L_{kl}]=0.
\end{eqnarray}

\bigskip
\bigskip
In fact, by considering the definition (\ref{omega}) and the relation (\ref{2nddef}), from the coefficients of $d \lambda \wedge d z$,$d \lambda \wedge d t_{mn}$
and $d z \wedge d t_{mn}$ respectively, one obtains
\begin{subequations}
\begin{eqnarray}
1&=&\frac{\partial \Phi}{\partial \lambda} \frac{\partial \Psi}{\partial z}-\frac{\partial \Phi}{\partial z} \frac{\partial \Psi}{\partial \lambda}=\{\Phi,\Psi\},
\end{eqnarray}
\begin{eqnarray}
\frac{\partial H_{mn}}{\partial \lambda}&=&\frac{\partial \Phi}{\partial \lambda} \frac{\partial \Psi}{\partial t_{mn}}-\frac{\partial \Phi}{\partial t_{mn}} \frac{\partial \Psi}{\partial \lambda},
\end{eqnarray}
\begin{eqnarray}
\frac{\partial H_{mn}}{\partial z}&=&\frac{\partial \Phi}{\partial z} \frac{\partial \Psi}{\partial t_{mn}}-\frac{\partial \Phi}{\partial t_{mn}} \frac{\partial \Psi}{\partial z}.
\end{eqnarray}
\end{subequations}
By solving these equations for ${\partial \Phi}/{\partial t_{mn}}$ and  ${\partial \Psi}/{\partial t_{mn}}$, one  obtains
\begin{subequations}
\begin{eqnarray}
\frac{\partial \Phi}{\partial t_{mn}}=\{H_{mn},\Phi\},
\end{eqnarray}
\begin{eqnarray}
\frac{\partial \Psi}{\partial t_{mn}}=\{H_{mn},\Psi\},
\end{eqnarray}
\end{subequations}
which are equivalent to the following expressions
\begin{subequations}
\begin{eqnarray}
L_{mn} \Phi=0,
\end{eqnarray}
\begin{eqnarray}
L_{mn} \Psi=0,
\end{eqnarray}
\end{subequations}
Analogously, we have the following results for the $\hat{\Phi}$ and $\hat{\Psi}$
\begin{subequations}
\begin{eqnarray}
\frac{\partial \hat{\Phi}}{\partial t_{mn}}=\{H_{mn},\hat{\Phi}\},
\end{eqnarray}
\begin{eqnarray}
\frac{\partial \hat{\Psi}}{\partial t_{mn}}=\{H_{mn},\hat{\Psi}\},
\end{eqnarray}
\end{subequations}
which can be expressed as
\begin{subequations}
\begin{eqnarray}
L_{mn} \hat{\Phi}=0,
\end{eqnarray}
\begin{eqnarray}
L_{mn} \hat{\Psi}=0.
\end{eqnarray}
\end{subequations}

\bigskip
Some examples from the dDS hierarchy are presented below.

\bigskip
\textbf{Example 1:}
By taking $m=2,n=0;k=3,l=0$ and $t_{20}=y,t_{30}=t$, Hamiltonians read as follows
\begin{subequations}
\begin{eqnarray}
H_{20}&=&\mathcal{A}_{2}=\lambda^{2}+2 f_{0} \lambda \nonumber\\
&\equiv& \lambda^{2}+2 f \lambda ,
\end{eqnarray}
\begin{eqnarray}
H_{30}&=&\mathcal{A}_{3}=\lambda^{3}+3 f_{0} \lambda^{2}+(3f_{-1}+3f^{2}_{0})\lambda  \nonumber\\
&\equiv&\lambda^{3}+3 f \lambda^{2}+w\lambda.
\end{eqnarray}
\end{subequations}
Then the following nonlinear system
\begin{subequations}
\begin{eqnarray}
3f_{y}-2w_{z}+6ff_{z}=0,
\end{eqnarray}
\begin{eqnarray}
2f_{t}-w_{y}+2fw_{z}-2f_{z}w=0,
\end{eqnarray}
\end{subequations}

arises from the Zakharov-Shabat equation (\ref{zerocurvature}) which is nothing but the complex dmKP (dispersionless modified Kadomtsev-Petviashvili) system.

\bigskip
\textbf{Example 2:}
By taking $m=1,n=1;k=2,l=2$, Hamiltonians read as follows
\begin{subequations}
\begin{eqnarray}
H_{11}= \lambda+(v+\frac{u}{\lambda}),
\end{eqnarray}
\begin{eqnarray}
H_{22}&=&(\lambda^{2}+2f_{0} \lambda)+ [(v^{2}+2u g_{1})+\frac{2uv}{\lambda}+\frac{u^{2}}{\lambda^{2}}] \nonumber\\
&\equiv& (\lambda^{2}+2 f \lambda)+ [w+\frac{2uv}{\lambda}+\frac{u^{2}}{\lambda^{2}}].
\end{eqnarray}
\end{subequations}
Then the following nonlinear system
\begin{subequations}
\begin{eqnarray}
2u_{t_{11}}-2u_{z}+w_{z}-2vv_{z}=0,
\end{eqnarray}
\begin{eqnarray}
u_{t_{22}}+2(uv)_{z}-2(uv)_{t_{11}}-2(uf)_{z}=0,
\end{eqnarray}
\begin{eqnarray}
v_{t_{22}}-w_{t_{11}}+w_{z}-2v_{z}f-2u_{z}=0,
\end{eqnarray}
\begin{eqnarray}f_{t_{11}}-f_{z}+v_{z}=0.
\end{eqnarray}
\end{subequations}
arises from the Zakharov-Shabat equation (\ref{zerocurvature}).

\textbf{Example 3:}
By taking $m=0,n=1;k=2,l=2$ and $t_{01}=\hat{z},t_{22}=t$, Hamiltonians read as follows
\begin{subequations}
\begin{eqnarray}
H_{01}= v+\frac{u}{\lambda},
\end{eqnarray}
\begin{eqnarray}
H_{22}&=&(\lambda^{2}+2f_{0} \lambda)+ [(v^{2}+2u g_{1})+\frac{2uv}{\lambda}+\frac{u^{2}}{\lambda^{2}}] \nonumber\\
&\equiv& (\lambda^{2}+2 f \lambda)+ [w+\frac{2uv}{\lambda}+\frac{u^{2}}{\lambda^{2}}].
\end{eqnarray}
\end{subequations}
Then nonlinear system arise from the Zakharov-Shabat equation  (\ref{zerocurvature}) read as
\begin{subequations}
\begin{eqnarray}
2u_{\hat{z}}+w_{z}-2vv_{z}=0,
\end{eqnarray}
\begin{eqnarray}
u_{t}-2(uf)_{z}-2(uv)_{\hat{z}}=0,
\end{eqnarray}
\begin{eqnarray}
v_{t}-2u_{z}-2fv_{z}- w_{\hat{z}}=0,
\end{eqnarray}
\begin{eqnarray}
f_{\hat{z}}+v_{z}=0.
\end{eqnarray}
\end{subequations}
which can be simplified exactly to the  dDS system (\ref{dDS system}) with the choice $v=S_{\hat{z}}$.

\textbf{Example 4:} If one of the two Hamiltonians is chosen as $\hat{H}=H_{01}$, by virtue of (\ref{simplezerocurvature})and (\ref{simplesystems}), the integrable flow equations can be obtained through (\ref{simplesystems}) simply. For example, by choosing another Hamiltonian as $H_{33}$ and letting $v=S_{\hat{z}}$,
the functions $f^{(3)}_{0},f^{(3)}_{1},g^{(3)}_{0}$ and $g^{(3)}_{-1}$ in the expression (\ref{simplesystems}) read as follows
\begin{subequations}
\begin{eqnarray}
f^{(3)}_{0}=-S^{3}_{z}+6S_{z} V+\partial^{-1}_{\hat{z}} (u S_{zz}-S_{z\hat{z}} V),
\end{eqnarray}
\begin{eqnarray}
f^{(3)}_{1}=3(S^{2}_{z}-V),
\end{eqnarray}
\begin{eqnarray}
g^{(3)}_{0}=S^{3}_{\hat{z}}-6S_{\hat{z}} W+\partial^{-1}_{z} (S_{z\hat{z}} W-u S_{\hat{z}\hat{z}}),
\end{eqnarray}
\begin{eqnarray}
g^{(3)}_{-1}=3u(S^{2}_{\hat{z}}-W),
\end{eqnarray}
\end{subequations}
in which
\begin{eqnarray}
W_{z}=u_{\hat{z}},~~~~~~V_{\hat{z}}=u_{z}.
\end{eqnarray}
Then one obtains the following system
\begin{subequations}
\begin{eqnarray}
u_{t_{33}}-3\left[u(S^{2}_{z}-V)\right]_{z}-3\left[u(S^{2}_{\hat{z}}-W)\right]_{\hat{z}}=0,
\end{eqnarray}
\begin{eqnarray}
S_{t_{33}}-(S^{3}_{z}+S^{3}_{\hat{z}})+3(S_{z} V+S_{\hat{z}} W)+3\phi=0,
\end{eqnarray}
\begin{eqnarray}
\phi_{z\hat{z}}=(uS_{z})_{zz}+(uS_{\hat{z}})_{\hat{z}\hat{z}},
\end{eqnarray}
\end{subequations}
which is analogous to the dDS system (\ref{dDS system}).

\bigskip
\textbf{Remark 3:} The dDS hierarchy generalizes the complex dmKP (dispersionless modified KP) hierarchy. In fact,
the complex dmKP hierarchy is the particular case for $n=l=0$ in (\ref{zerocurvature}), i.e., defined as the following zero-curvature equations
\begin{eqnarray}
&&\frac{\partial H_{m0}}{\partial t_{k0}}-\frac{\partial H_{k0}}{\partial t_{m0}}+\{H_{m0},H_{k0}\} \nonumber\\
&=&\frac{\partial \mathcal{A}_{m}}{\partial t_{k0}}-\frac{\partial \mathcal{A}_{k}}{\partial t_{m0}}+\{\mathcal{A}_{m},\mathcal{A}_{k}\}=0.
\end{eqnarray}

\textbf{Remark 4:} The general unreduced  hierarchy (without the canonical Poisson relations (\ref{canonical}), similar to the Manakov-Santini hierarchy \cite{manakov2,manakov3}), can also be defined  by the following Lax-Sato equations
\begin{eqnarray}
\mathcal{L}_{mn} (K)=\frac{\partial K}{\partial t_{mn}}-P_{mn} \frac{\partial K}{\partial z}+S_{mn} \frac{\partial K}{\partial \lambda}=0
\end{eqnarray}
for eigenfunctions $K=\Phi,\Psi,\hat{\Phi},\hat{\Psi}$ defined in (\ref{eigen}), in which
\begin{subequations}
\begin{eqnarray}
P_{mn}=(m  \Phi^{m-1} \Phi_{\lambda} J^{-1})_{\geq 0}+(n \hat{\Phi}^{n-1} \hat{\Phi}_{\lambda}\hat{J}^{-1})_{< 0},
\end{eqnarray}
\begin{eqnarray}
S_{mn}=(m \Phi^{m-1} \Phi_{z} J^{-1})_{>0}+(n \hat{\Phi}^{n-1} \hat{\Phi}_{z}\hat{J}^{-1})_{\leq 0},
\end{eqnarray}
\end{subequations}
here $J=\{\Phi,\Psi\},\hat{J}=\{ \hat{\Phi}, \hat{\Psi}\}$ and $ J^{-1},\hat{J}^{-1}$ are their formal inverses respectively.
Obviously, this definition is equivalent to
\begin{eqnarray}
[\mathcal{L}_{mn},\mathcal{L}_{kl}]=0.
\end{eqnarray}

\bigskip
\bigskip
\textbf{Acknowledgements:} This work has been supported by the National Natural Science Foundation of China (No. 11501222). The author would like to thank Professor Paolo Maria Santini very sincerely for the valuable suggestions and patient guidance to accomplishment of this work. The author also would like to thank Professor Maciej Dunajski for the valuable comments and suggestions via email.

\end{document}